\documentclass[twocolumn,showpacs,preprintnumbers,amsmath,amssymb]{revtex4}

\usepackage{graphicx}
\usepackage{dcolumn}
\usepackage{bm}

\begin{document}

\title{Coherent population trapping and dynamical instability in the
nonlinearly coupled atom-molecule system}
\author{H. Y. Ling and P. Maenner }
\affiliation{Department of Physics and Astronomy, Rowan University, Glassboro, NJ
08028-1700}
\author{H. Pu }
\affiliation{Department of Physics and Astronomy, and Rice Quantum Institute, Rice
University, Houston, TX 77251-1892}

\begin{abstract}
We study the possibility of creating a coherent population
trapping (CPT) state, involving free atomic and ground molecular
condensates, during the process of associating atomic condensate
into molecular condensate. We generalize the Bogoliubov approach
to this multi-component system and study the collective
excitations of the CPT state in the homogeneous limit. We develop
a set of analytical criteria based on the relationship among
collisions involving atoms and ground molecules, which are found
to strongly affect the stability properties of the CPT state, and
use it to find the stability diagram and to systematically
classify various instabilities in the long-wavelength limit.
\end{abstract}

\date{\today }
\maketitle

\section{Introduction}

The availability of atomic Bose-Einstein condensates (BECs) has
made it possible to create, via photo- or magneto-association,
molecular condensates. In photoassociation, a pair of free atoms
is brought into a bound molecular state by absorption of a photon
through a dipole transition. In magneto-association (or Feshbach
resonance \cite{Feshbach,timmermans99}), atom pairs in an open channel are
converted into bound molecules in a closed channel through
hyperfine spin interaction resonantly enhanced by making the
energies of the two channels close to each other via the Zeeman
effect. Photoassociation creates molecules in excited electronic
level, while magneto-association creates molecules in high
vibrational quantum state. In either case, the resulting molecules
are energetically unstable and suffer from large inelastic loss
rate. Macroscopic coherence between atoms and molecules has been
both studied theoretically \cite%
{timmermans99,drummond98,javanainen99,yurovsky99,heinzen00,kokkelmans02,javanainen02} (also see a recent review article by Duine and Stoof \cite{duine04}), and
demonstrated experimentally \cite{donley02}, nevertheless, long-lived stable molecular
condensates have not been produced by direct association of atomic
condensate \cite{fbecnote}.

\begin{figure}[h]
\centering \includegraphics[width=2.in]{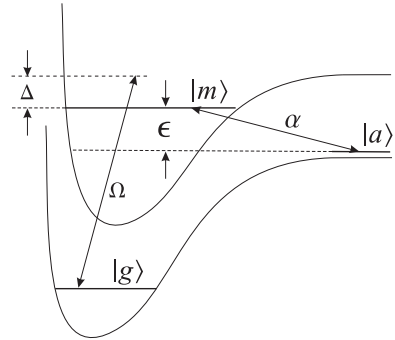} \caption{The
energy diagrams of three-level atom-molecule system involving
free-bound-bound transitions. Conversion of atoms in $\left\vert
a\right\rangle $ to quasi-bound molecules in $\left\vert
m\right\rangle $ is accomplished by Feshbach resonance, while the
coupling between $|m \rangle$ and the ground molecular state
$|g\rangle$ is provided by laser light.} \label{fig1}
\end{figure}

One possibility to overcome the difficulty is to employ the
two-photon or Raman photoassociation model, as in the proposal for
generating ground molecular condensates \cite{thermal,Mackie00,hope01,Drummond02}
from atomic condensates through the stimulated Raman adiabatic
passage (STIRAP) \cite{Hioe83,stirap}. In spite of the nonlinear
nature of the atom-molecule coupling in the photoassociation
process, it was shown by Mackie \emph{et al}. \cite{Mackie00} that
in the collisionless limit, this model can support a coherent
superposition of free atomic and ground molecular condensates.
This superposition, being immune to the inelastic processes
associated with the excited molecular state, is called the
nonlinear coherent population trapping (CPT) state, in direct
analogy to the CPT state in the linear $\Lambda $ three-level
atomic system \cite{Alzetta76,Gray78}. However, efficient
operation of this scheme typically demands very high laser
intensities in order to overcome the weak coupling caused by the
extremely small Franck-Condon overlap integral for the free
atomic-bound molecular transition \cite{Drummond02}.

To overcome the small Franck-Condon factor, a Feshbach-assisted
STIRAP model is proposed \cite{kokk,mackie}. Here, conversion of
atoms into quasi-bound molecules is accomplished by the much more
efficient Feshbach process. To bring the quasi-bound molecules
into a more deeply bound vibrational state, a two-photon Raman
laser field is applied. In the limit where both optical fields are
tuned far-off resonance from other electronically excited levels,
one can reduce the Raman coupling between the two molecular levels
into an effective one-photon coupling. This leads to a three-level
$\Lambda $ atom-molecule system \cite{Harris02} as shown in
Fig.~\ref{fig1}. In this scheme, the possible source of
decoherence is the decay of the quasi-bound molecular state while
the atomic state and the deeply bound molecular state are both
stable.

In the absence of particle collisions, we expect, from the work of
Mackie \emph{et al}. \cite{Mackie00}, that this system is capable
of supporting a CPT state, owing to the mathematical equivalence
to its photoassociation counterpart. The situation when the
collisions are present is more complicated. It was generally
believed that the nonlinear phase shifts arising from particle
collisions would defeat the two-photon resonance condition
\cite{Mackie00,Drummond02}, a prerequisite for the existence of
CPT states and hence the key for the successful implementation of
STIRAP. In a recent work \cite{ling04}, we have shown that by
appropriately chirping the optical frequency and the Feshbach
detuning, the collision-induced nonlinear phase shifts can be
dynamically compensated, hence generalizing the concept of the CPT
state from the collisionless limit \cite{Mackie00} to situations
where collisions cannot be ignored. In addition, we have also
shown that collisions may lead to dynamical instabilities in the
nonlinear CPT state, in contrast to the CPT states in the linear
$\Lambda $-system which are always stable. Avoiding dynamical
unstable regimes is another key for successful implementation of
the STIRAP process in nonlinear systems.

The purpose of the current paper is to provide a systematic study
of the excitation spectrum of the generalized nonlinear CPT state,
focusing on the characterization of its stability phase diagram.
To the best of our knowledge, a detailed stability analysis
specific to the CPT state of the nonlinear $\Lambda $ system has
never been presented before. Our paper is organized as follows: In
Sec.~II, we present the Hamiltonian and the nonlinear CPT solution
to the coupled Gross-Pitaevskii equations. In Sec.~III, collective
excitations of the CPT state are calculated in the spirit of the
Bogoliubov treatment \cite{bogoliubov47,fetter71}. Sec.~IV focuses
on the stability analysis in the long-wavelength limit based on
the excitation spectrum obtained in Sec.~III. In Sec. V, we study
the dynamics of a CPT state and show that dynamical instabilities
may lead to self-pulsing of a condensate mode or growth of
non-condensate modes. Finally, a summary is provided in Sec. VI.

\section{Hamiltonian, Stationary Equations, and CPT State}

We choose to formulate our theoretical description for the
Feshbach-based $\Lambda $-system as illustrated in
Fig.~\ref{fig1}. The exactly same description can be applied to
the two-photon photoassociation-based $\Lambda $-system, due to
the formal equivalence between the two models \cite{Drummond02}.
In Fig.~\ref{fig1}, we use $|a\rangle $, $|m\rangle $, and
$|g\rangle $ to denote the free atomic, quasi-bound molecular and
ground molecular states, respectively. Levels $|a\rangle $ and
$|m\rangle $ are coupled by a magnetic field through the Feshbach
resonance with a coupling strength $\alpha $ and a detuning
$\epsilon $ ($\epsilon $ is experimentally tunable via the
magnetic field), while levels $|m\rangle $ and $|g\rangle $ are
coupled by a (effective) laser field with a Rabi frequency $\Omega
$ and a detuning $\Delta$. In second quantization, the total
Hamiltonian, including collisions, takes the form
\begin{widetext}
\begin{eqnarray}
\hat{H} &=&\hbar \int d\mathbf{r} \left\{
\sum_{i}\hat{\psi}_{i}^{\dag }\left( \mathbf{r}\right) \left(
-\frac{\hbar }{2m_{i}}\nabla ^{2}\right) \hat{\psi}_{i}\left(
\mathbf{r}\right) +\frac{1}{2}\sum_{i,j}\lambda _{ij}
\hat{\psi}_{i}^{\dag }\left( \mathbf{r}\right)
\hat{\psi}_{j}^{\dag }\left( \mathbf{r}\right)
\hat{\psi}_{j}\left( \mathbf{r}\right) \hat{\psi}
_{i}\left( \mathbf{r}\right) \right. \nonumber \\
&&+\left.\epsilon \, \hat{\psi}_{m}^{\dag }\left(
\mathbf{r}\right) \hat{\psi} _{m}\left( \mathbf{r}\right)
+\frac{\alpha }{2}\left[ \hat{\psi}_{m}^{\dag }\left(
\mathbf{r}\right) \hat{\psi}_{a}\left( \mathbf{r}\right)
\hat{\psi} _{a}\left( \mathbf{r}\right) +h.c.\right] +\left(
\Delta +\epsilon \right) \hat{\psi}_{g}^{\dag }\left(
\mathbf{r}\right) \hat{\psi}_{g}\left( \mathbf{r}\right)
-\frac{\Omega }{2} \left[ \hat{\psi}_{m}^{\dag }\left(
\mathbf{r}\right) \hat{\psi}_{g}\left( \mathbf{r}\right) +h.c.
\right] \right\}, \label{grandH}
\end{eqnarray}
\end{widetext}
where $\hat{\psi}_{i}^{\dag }\left( \mathbf{r}\right) $ [$\hat{
\psi}_{i}\left( \mathbf{r}\right) $] ($i=a$, $m$ and $g$) is the
creation
(annihilation) operator of the bosonic field for species $i$ at location $%
\mathbf{r}$,$\ $and the terms proportional to $\lambda _{ij}$ represent
two-body collisions with $\lambda _{ii}$ $=4\pi \hbar a_{i}/m_{i}$ and $%
\lambda _{ij}=\lambda _{ji}=2\pi \hbar a_{ij}/\mu _{ij}$ for $i\neq j\,$\
characterizing the intra- and inter-state interaction strengths,
respectively, ($a_{i}$ and $a_{ij}$ are $s$-wave scattering lengths, $%
m_{i} $ is the mass of species $i$ with $m_{m}=m_{g}=2m_{a}$, and
$\mu _{ij}$ is the reduced mass between states $i$ and $j$). We
take, without loss of generality, both $\alpha $ and $\Omega $ to
be real as their phase factors can be absorbed by a trivial global
gauge transformation of the field operators. Here, we consider a
uniform system and hence have dropped the external trapping
potentials.

Let us begin our formulation from the Bogoliubov approximation,
which amounts to decomposing $\hat{\psi}_{i}\left(
\mathbf{r}\right) $ as
\begin{equation}
\hat{\psi}_{i}\left( \mathbf{r}\right) \approx \psi _{i}\left( \mathbf{r}%
\right) +\delta \hat{\psi}_{i}\left( \mathbf{r}\right) ,  \label{bogoliubov}
\end{equation}%
where $\psi _{i}\left( \mathbf{r}\right) =\left\langle
\hat{\psi}_{i}\left( \mathbf{r}\right) \right\rangle $ is the
condensate wave function and $\delta \hat{\psi}_{i}\left(
\mathbf{r}\right) $ is a small fluctuation field operator that
obeys the usual bosonic commutation relation
\begin{equation}
\left[ \delta \hat{\psi}_{i}\left( \mathbf{r}\right) ,\delta \hat{\psi}%
_{j}^{\dag }\left( \mathbf{r}^{\prime }\right) \right] =\delta _{ij}\delta
\left( \mathbf{r-r}^{\prime }\right) .  \label{space
communtator relation}
\end{equation}%
As a first step in the Bogoliubov approach, we put
Eq.~(\ref{bogoliubov}) into the grand canonical Hamiltonian
(\ref{grandH})
\begin{equation}
\mathcal{\hat{H}}=\hat{H}-\hbar \mu \int d\mathbf{r}\left[ \hat{\psi}%
_{a}^{\dag }\hat{\psi}_{a}+2\hat{\psi}_{m}^{\dag }\hat{\psi}_{m}+2\hat{\psi}%
_{g}^{\dag }\hat{\psi}_{g}\right] ,  \label{grand grand H}
\end{equation}%
where the chemical potential $\hbar \mu $ is introduced to
conserve the total particle number, and expand Eq.~(\ref{grand
grand H}) up to the second order in fluctuation operators as
\begin{equation}
\mathcal{\hat{H}}\approx \mathcal{\hat{H}}^{\left( 0\right) }+\mathcal{\hat{H%
}}^{\left( 1\right) }+\mathcal{\hat{H}}^{\left( 2\right) },  \label{H 012}
\end{equation}%
where the subscript represents the order in $\delta \hat{\psi}_{i}\left(
\mathbf{r}\right) $.

The zeroth order term, $\mathcal{\hat{H}}^{\left( 0\right) }$, is a constant
depending on the wave functions $\psi_i$ and is irrelevant to our interest.
The first-order term, $\mathcal{\hat{H}}^{\left( 1\right) }$, is required to
vanish in the Bogoliubov formalism. This requirement leads to a set of
coupled stationary equations
\begin{subequations}
\label{steady state equation}
\begin{eqnarray}
\mu \psi _{a}&=& \left( -\frac{\hbar \nabla ^{2}}{2m_{a}} +\sum_{i}\lambda
_{ai}\left\vert \psi _{i}\right\vert ^{2}\right) \psi _{a}+\alpha \psi
_{m}\psi _{a}^{\ast } ,  \label{u equation} \\
2\mu \psi _{m} &=& \left( -\frac{\hbar \nabla ^{2}}{4m_{a}} +\sum_{i}\lambda
_{mi}\left\vert \psi _{i}\right\vert ^{2}\right) \psi _{m}+\epsilon \psi
_{m}+\frac{\alpha }{2}\psi _{a}^{2}  \notag \\
&&\;\;\;\;\;\;-\frac{\Omega }{2}\psi _{g} ,  \label{amplitude} \\
2\mu \psi _{g} &=& \left( -\frac{\hbar \nabla ^{2}}{4m_{a}} +\sum_{i}\lambda
_{gi}\left\vert \psi _{i}\right\vert ^{2}\right) \psi _{g}+\left( \Delta
+\epsilon \right) \psi _{g}  \notag \\
&&\;\;\;\;\;\;-\frac{\Omega }{2}\psi _{m} ,  \label{delta equation}
\end{eqnarray}
which represent the generalized time-independent Gross-Pitaevskii
equations (GPEs). Finally, the second-order term,
$\mathcal{\hat{H}}^{\left( 2\right) } $, which has a more
complicated structure, will be studied separately in the next
section where we calculate the spectra of the collective
excitations.

We now turn our attention to the homogeneous (zero-momentum) solution to the
GPEs. To proceed, we separate each wave function into an amplitude and a
phase according to
\end{subequations}
\begin{equation}
\psi _{j}=\left\vert \psi _{j}\right\vert e^{i\theta _{j}},\;\;j=a,m,g.
\label{steady state ansatz}
\end{equation}%
To seek the CPT solution, we take $\psi _{m}=0$. After
substituting Eq.~(\ref{steady state ansatz}) into Eq.~(\ref{steady
state equation}), we find that under the conditions
\begin{eqnarray}
\theta _{g} &=&2\theta _{a},  \notag \\
\alpha \left\vert \psi _{a}\right\vert ^{2} &=&\Omega \left\vert \psi
_{g}\right\vert ,  \label{amplitude relation}
\end{eqnarray}%
Eq.~(\ref{steady state equation}) possesses a solution consistent
with the requirement that $\psi _{m}=0$. The explicit form of this
solution can be readily found by combining Eq.~(\ref{amplitude
relation} ) with the particle number conservation
\begin{equation}
\left\vert \psi _{a}\right\vert ^{2}+2\left\vert \psi _{g}\right\vert ^{2}=n,
\label{particle number conservation}
\end{equation}%
where $n$ is the total particle density, to yield the particle
density distribution as
\begin{subequations}
\label{CPT distribution}
\begin{eqnarray}
\left\vert \psi _{m}^{0}\right\vert ^{2} &=&0\,, \\
\left\vert \psi _{a}^{0}\right\vert ^{2} &=&\frac{2n}{1+\sqrt{1+8\left(
\frac{\alpha \sqrt{n}}{\Omega }\right) ^{2}}}\,, \\
\left\vert \psi _{g}^{0}\right\vert ^{2} &=&\frac{n}{2}\frac{\sqrt{1+8\left(
\frac{\alpha \sqrt{n}}{\Omega }\right) ^{2}}-1}{1+\sqrt{1+8\left( \frac{%
\alpha \sqrt{n}}{\Omega }\right) ^{2}}}\,.
\end{eqnarray}%
A check for consistency leads to the chemical potential
\end{subequations}
\begin{equation}
\mu =\lambda _{a}\left\vert \psi _{a}^{0}\right\vert ^{2}+\lambda
_{ag}\left\vert \psi _{g}^{0}\right\vert ^{2},  \label{chemical potential}
\end{equation}%
where $\lambda _{i}\equiv \lambda _{ii}$, and the laser detuning
\begin{equation}
\Delta =-\epsilon +\left( 2\lambda _{ag}-\lambda _{g}\right) \left\vert \psi
_{g}^{0}\right\vert ^{2}+\left( 2\lambda _{a}-\lambda _{ag}\right)
\left\vert \psi _{a}^{0}\right\vert ^{2},  \label{Delta arbitrary}
\end{equation}%
which is necessary in order for Eqs.~(\ref{CPT distribution}) to be a
stationary solution of Eqs.~(\ref{steady state equation}).

The vanishing of the population in the quasi-bound molecular state
indicates a destructive interference in the excitation to this
state. A system prepared in the state represented by
Eqs.~(\ref{CPT distribution}) is not subject to the particle loss
suffered by the quasi-bound molecular state. This situation is
reminiscent of the CPT state in a linear $\Lambda $-type atomic
system where the atoms are immune to the spontaneous emission
\cite{Alzetta76,Gray78}. For this reason, we regard the steady
state represented by Eqs.~(\ref{CPT distribution}) as the
nonlinear matter-wave analog of the CPT state in a linear $\Lambda
$-system.

Interestingly, just as in the linear case, the CPT solution
described by Eqs.~(\ref{CPT distribution}) depends only explicitly
on the two coupling strengths, $\alpha $ and $\Omega $, not on the
collisional parameters. The latter, however, play important roles
as evidenced by Eq.~(\ref{Delta arbitrary}). Note that in the
absence of collisions, Eq.~(\ref{Delta arbitrary}) degenerates
into $\Delta +\epsilon =0$, which is just the ordinary
\textquotedblleft two-photon\textquotedblright\ resonance
condition. Thus, Eq.~(\ref{Delta arbitrary}) can be regarded as
the generalization of the two-photon resonance in which the
nonlinear phase shifts due to particle collisions have been
compensated. This CPT solution is therefore a nonlinear
generalization of the one found in the collisionless limit
\cite{Mackie00}.

\section{Collective Excitations of the CPT State}

In this section, we want to calculate the excitation spectra of
the CPT state, especially, the spectra in the long-wavelength
limit, since they determine many important properties, including
the stability, of the CPT state at low temperature.

\subsection{Bogoliubov Equations}

To determine the equations for the collective excitations, we
first take advantage of our system being homogeneous and move to
the momentum space through the expansion
\begin{equation}
\delta \hat{\psi}_{i}\left( \mathbf{r}\right) =\frac{1}{\left( 2\pi \hbar
\right) ^{3/2}}\int d\mathbf{p}\,e^{i\mathbf{p\cdot r/\hbar }}\,\hat{c}%
_{i}\left( \mathbf{p}\right)\,,
\end{equation}%
where $\hat{c}_{i}\left( \mathbf{p}\right) $ [$\hat{c}_{i}^{\dag }\left(
\mathbf{p}\right) $] is the annihilation (creation) operator for a particle
of species $i$ with momentum $\mathbf{p}$ and obey the standard bosonic
commutation relation
\begin{equation}
\left[ \hat{c}_{i}\left( \mathbf{p}\right) ,\hat{c}_{j}^{\dag }\left(
\mathbf{p}^{\prime }\right) \right] =\delta _{ij}\delta \left( \mathbf{p-p}%
^{\prime }\right) ,  \label{c bosonic commutation}
\end{equation}%
in accordance with Eq.~(\ref{space communtator relation}). The third term $%
\mathcal{\hat{H}}^{\left( 2\right) }$ in Eq.~(\ref{H 012}) for our CPT state
can now be put into a compact form
\begin{eqnarray}
\mathcal{\hat{H}}^{\left( 2\right) } &=&\frac{\hbar }{2}\int d\mathbf{p}%
^{\prime }\sum_{nm}\left\{ 2\hat{c}_{n}^{\dag }\left( \mathbf{p}^{\prime
}\right) \left( A\right) _{nm}\hat{c}_{m}\left( \mathbf{p}^{\prime }\right)
\right.  \notag \\
&&+\left. \left[ \hat{c}_{n}\left( \mathbf{p}^{\prime }\right) B_{nm}\hat{c}%
_{m}\left( -\mathbf{p}^{\prime }\right) +h.c.\right] \right\} ,  \label{H^2}
\end{eqnarray}%
where $A$ and $B$ are both real and symmetric matrices, given by
\begin{subequations}
\begin{eqnarray}
A &=&\left(
\begin{array}{ccc}
\epsilon _{p}+\lambda _{a}\left\vert \psi _{a}^{0}\right\vert ^{2} & \alpha
\left\vert \psi _{a}^{0}\right\vert & \lambda _{ag}\left\vert \psi
_{a}^{0}\right\vert \left\vert \psi _{g}^{0}\right\vert \\
\alpha \left\vert \psi _{a}^{0}\right\vert & \frac{\epsilon _{p}}{2}%
+\epsilon ^{\prime } & -\frac{\Omega }{2} \\
\lambda _{ag}\left\vert \psi _{a}^{0}\right\vert \left\vert \psi
_{g}^{0}\right\vert & -\frac{\Omega }{2} & \frac{\epsilon _{p}}{2}+\lambda
_{g}\left\vert \psi _{g}^{0}\right\vert ^{2}%
\end{array}%
\right) , \\
B &=&\left(
\begin{array}{ccc}
\lambda _{a}\left\vert \psi _{a}^{0}\right\vert ^{2} & 0 & \lambda
_{ag}\left\vert \psi _{a}^{0}\right\vert \left\vert \psi _{g}^{0}\right\vert
\\
0 & 0 & 0 \\
\lambda _{ag}\left\vert \psi _{a}^{0}\right\vert \left\vert \psi
_{g}^{0}\right\vert & 0 & \lambda _{g}\left\vert \psi _{g}^{0}\right\vert
^{2}%
\end{array}%
\right) ,
\end{eqnarray}%
\end{subequations}
where we have defined
\begin{subequations}
\begin{eqnarray}
\epsilon _{p} &\equiv &p^{2}/(2m_{a}\hbar )\,,   \\
\epsilon ^{\prime } &=&\epsilon -\Lambda _{a}\left\vert \psi
_{a}^{0}\right\vert ^{2}-\Lambda _{g}\left\vert \psi _{g}^{0}\right\vert
^{2}\,,  \label{shifted epsilon} \\
\Lambda _{a} &=&2\lambda _{a}-\lambda _{am}\,,  \\
\Lambda _{g} &=&2\lambda _{ag}-\lambda _{mg}\,.
\end{eqnarray}
\end{subequations}

The process of finding the quasiparticle spectrum begins by changing Eq.~(%
\ref{H^2}) into a diagonalized form
\begin{equation}
\mathcal{\hat{H}}^{\left( 2\right) }=\hbar \int d\mathbf{p}\sum_{i}\omega
_{i}\left( p\right) \hat{b}_{i}^{\dag }\left( \mathbf{p}\right) \hat{b}%
_{i}\left( \mathbf{p}\right) + c{\rm -number},
\end{equation}%
with the generalized Bogoliubov transformation
\begin{equation}
\hat{b}_{i}\left( \mathbf{p}\right) =\sum_{j}u_{ij}\left( p\right) \hat{c}%
_{j}\left( \mathbf{p}\right) +\sum_{j}v_{ij}\left( p\right) \hat{c}%
_{j}^{\dag }\left( -\mathbf{p}\right) ,  \label{bogoliubov transformation}
\end{equation}%
where $\omega _{i}\left( p\right) $ is the quasiparticle frequency, $%
u_{ij}\left( p\right) $ and $v_{ij}\left( p\right) $ are the transformation
coefficients, and $\hat{b}_{i}\left( \mathbf{p}\right) $ and $\hat{b}%
_{i}^{\dag }\left( \mathbf{p}\right) $ are the respective annihilation and
creation operators for a quasiparticle of frequency $\omega _{i}\left(
p\right) $ and momentum $\mathbf{p}$. To obtain the equations for $\omega
_{i}\left( p\right) $, $u_{ij}\left( p\right) $ and $v_{ij}\left( p\right) $%
, we first note that
\begin{equation}
\left[ \hat{b}_{i}\left( \mathbf{p}\right) ,\mathcal{\hat{H}}^{\left(
2\right) }\right] =\hbar \omega _{i}\left( p\right) \hat{b}_{i}\left( \mathbf{p}%
\right) .  \label{operator equation}
\end{equation}%
Inserting Eq.~(\ref{bogoliubov transformation}) into Eq.~(\ref{operator
equation}) and with the help of Eqs.~(\ref{c bosonic commutation}) and (\ref%
{H^2}) along with the symmetric properties of $A$ and $B$, we arrive at
\begin{widetext}
\begin{equation}
\sum_{m}\sum_{n}\left\{ \left[ A_{mn}u_{in}-B_{mn}v_{in}\right] \hat{c}%
_{m}\left( \mathbf{p}\right) +\left[ B_{mn}u_{in}-A_{mn}v_{in}\right] \hat{c}%
_{m}^{\dag }\left( -\mathbf{p}\right) \right\} =\omega _{i}\sum_{m}\left[
u_{im}\hat{c}_{m}\left( \mathbf{p}\right) +v_{im}\hat{c}_{m}^{\dag }\left( -%
\mathbf{p}\right) \right] . 
\end{equation}%
\end{widetext}Equating the coefficients of $\hat{c}_{m}\left( \mathbf{p}%
\right) $ and $\hat{c}_{m}^{\dag }\left( -\mathbf{p}\right) $ on both sides
of the above equation, we arrive at a 6$\times 6$ matrix equation
\begin{equation}
\left(
\begin{array}{cc}
A & -B \\
B & -A%
\end{array}%
\right) \left(
\begin{array}{c}
u_{i} \\
v_{i}%
\end{array}%
\right) =\omega _{i}\left( p\right) \left(
\begin{array}{c}
u_{i} \\
v_{i}%
\end{array}%
\right) ,  \label{collective excitation}
\end{equation}%
where
\begin{equation*}
u_{i}=\left( u_{ia},u_{im},u_{ig}\right) ^{T},\;\;\;v_{i}=\left(
v_{ia},v_{im},v_{ig}\right) ^{T}.
\end{equation*}

A simple manipulation can show that $\omega_i^2(p)$ are the
eigenvalues of the $3 \times 3$ matrix $(A+B)(A-B)$, hence
$\omega_i^2(p)$ satisfy a cubic equation with the form
\begin{equation}
\left( \omega _{i}^{2}\right) ^{3}-a_{1}\left( \omega
_{i}^{2}\right) ^{2}+a_{2}\omega _{i}^{2}-a_{3}=0\,,
\label{eigenvalue equation}
\end{equation}%
where the coefficients are given by
\begin{subequations}
\begin{eqnarray}
a_{1} &=&a_{10}+a_{11}\epsilon _{p}+a_{12}\epsilon _{p}^{2}\,, \\
a_{2} &=&a_{20}+a_{21}\epsilon _{p}+a_{22}\epsilon _{p}^{2}+a_{23}\epsilon
_{p}^{3}+a_{24}\epsilon _{p}^{4}\,, \\
a_{3} &=&a_{31}\epsilon _{p}+a_{32}\epsilon _{p}^{2}+a_{33}\epsilon
_{p}^{3}+a_{34}\epsilon _{p}^{4}+  \notag \\
&&a_{35}\epsilon _{p}^{5}+a_{36}\epsilon _{p}^{6}\,.
\end{eqnarray}
\end{subequations}
The coefficients $a_{ij}$ depend on the coupling and the collisional
constants and are in general quite complicated. We only list three of them
below
\begin{widetext}
\begin{subequations}
\begin{eqnarray}
a_{10} &=&\epsilon ^{\prime 2}+\frac{1}{2}\left( \Omega ^{2}+4\alpha
^{2}\left\vert \psi _{a}^{0}\right\vert ^{2}\right)\, ,  \label{a10} \\
a_{20} &=&\frac{1}{16}\left( \Omega ^{2}+4\alpha ^{2}\left\vert \psi
_{a}^{0}\right\vert ^{2}\right) ^{2}-0.5\epsilon ^{\prime }\alpha
^{2}\left\vert \psi _{a}^{0}\right\vert ^{4}\left( \lambda _{g}-4\lambda
_{ag}+4\lambda _{a}\right)\, ,  \label{a20} \\
a_{31} &=&\frac{1}{8}\left\vert \psi _{a}^{0}\right\vert ^{2}\left( \Omega
^{2}+2\alpha ^{2}\left\vert \psi _{a}^{0}\right\vert ^{2}\right) \times %
\left[ \lambda _{a}\Omega ^{2}+4\alpha ^{2}\left( \lambda
_{g}\left\vert \psi _{g}^{0}\right\vert ^{2}+\lambda
_{ag}\left\vert \psi _{a}^{0}\right\vert ^{2}\right) -8\epsilon
^{\prime }\left( \lambda _{a}\lambda _{g}-\lambda _{ag}^{2}\right)
\left\vert \psi _{g}^{0}\right\vert ^{2}\right] \,, \label{a31}
\end{eqnarray}
\end{subequations}
\end{widetext}as they will be explicitly referenced in later discussions.

\subsection{Excitation Spectra}

Equation~(\ref{eigenvalue equation}) has a cubic form with respect to $%
\omega_i ^{2}$ and hence can be solved analytically. However the
analytical expressions are in general too complicated to provide
much physical insights. Since we are most interested in the
long-wavelength excitations, we proceed to take a
perturbative approach by expanding $\omega _{i}(p)$ in the limit of small $p$%
, or equivalently, small $\sqrt{\epsilon _{p}}$. The perturbative solution
thus takes the form:
\begin{equation}
\omega _{i}\left( p\right) \approx \omega _{i}^{\left( 0\right)
}+d_{i}^{\left( 1\right) }\,\sqrt{\epsilon _{p}}+d_{i}^{\left( 2\right)
}\,\epsilon _{p}+\cdots ,  \label{perturbative ansatz}
\end{equation}%
In order to bring out the effects of collisions more clearly, let
us first consider the collicionless limit.

\subsubsection{Excitation Spectra in the Collisionless Limit}

In the absence of collisions, the matrix $B$ becomes null. The excitation
frequencies $\omega _{i}(p)$ are eigenvalues of the Hermitian matrix $A$ and
therefore must be real. This means that there is no dynamical instability in
spite of the nonlinear nature of the coupling between the atomic and the
quasi-bound molecular state.

To the zeroth order, we find that $\omega _{i}^{\left( 0\right) }$ has three
solutions \cite{comment}
\begin{equation}
\omega _{0}^{\left( 0\right) }=0,\;\;\;\omega _{\pm }^{\left(
0\right) }= \sqrt{\left( a_{10}\pm
\sqrt{a_{10}^{2}-4a_{20}}\right) /2}\,, \label{omega (0)}
\end{equation}%
where the subscripts $\left( 0,-,+\right) $ are used to denote the
three branches of the collective modes. Clearly the 0-branch is
gapless, while the $ (\pm )$-branches are gapped. The reality of
$\omega _{\pm }^{\left( 0\right) }$ is guaranteed because here
both $a_{10}$ and $a_{20}$ are positive with $ a_{10}^{2}\geq
4a_{20}$ [see Eqs.~(\ref{a10}) and (\ref{a20}), in the absence of
collisions.]

For all three branches, we have found that $d_{i}^{(1)}=0$. Hence the next
leading order in $\omega _{i}(p)$ is quadratic in momentum with
\begin{subequations}
\begin{eqnarray}
d_{0}^{(2)} &=&\frac{\Omega ^{2}+2\alpha ^{2}\left| \psi _{a}^{0}\right| ^{2}%
}{\Omega ^{2}+4\alpha ^{2}\left| \psi _{a}^{0}\right| ^{2}}\,, \\
d_{\pm }^{\left( 2\right) } &=&\pm \frac{\left[ \frac{\Omega ^{2}}{8}+\alpha
^{2}\left| \psi _{a}^{0}\right| ^{2}+\frac{(\omega _{\pm }^{(0)})^{2}}{2}%
\right] }{\left[ (\omega _{\pm }^{(0)})^{2}+\frac{\Omega
^{2}}{4}+\alpha ^{2}\left| \psi _{a}^{0}\right| ^{2}\right] }\,,
\end{eqnarray}
\end{subequations}

\begin{figure}[h]
\centering \includegraphics[width=2.5in]{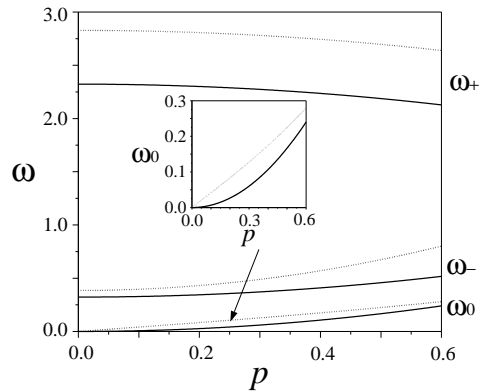} \caption{The
quasiparticle dispersion spectra for $n=1$, $\Omega =$ $1$, and
$\protect\epsilon =-2$. Solid lines represent the collisionless
case while dotted lines are for a system with $\protect
\lambda _{a}=0.625,$ $\protect\lambda _{m}=\protect\lambda _{g}=\protect%
\lambda _{am}=\protect\lambda _{ag}=\protect\lambda _{mg}=0.3\protect\lambda %
_{a}$. Units are defined in Sec.~\protect\ref{classification
section}.} \label{fig2}
\end{figure}
Solid lines in Fig.~\ref{fig2} represent the three branches in the
collisionless limit. They are obtained by directly solving the cubic
equation (\ref{eigenvalue equation}), and are in good agreement with the
perturbative results (not shown).

\subsubsection{Excitation Spectra with Collisions}

\label{spectra} The spectra including the collisional terms can be
similarly obtained. The zeroth order solutions, $\omega
_{i}^{(0)}$, have exactly the same forms as in the collisionless
limit given by Eq.~(\ref{omega (0)}). Therefore the 0-branch
continues to be gapless. Due to the collision-modified
coefficients $a_{ij}$, the two zeroth order gapped modes $\omega
_{\pm }^{(0)}$ are now, however, no longer guaranteed to be real.
Complex mode frequencies lead to the dynamical instability, which
we will discuss in detail in the next section.

Another important effect of the collisions is the modification of
the gapless 0-branch. While $d_{\pm }^{(1)}$ still remains to be
zero (hence the two gapped branches are still quadratic in $p$ to
the leading order), $ d_{0}^{(1)}$ now takes a finite value given
by
\begin{equation}
d_{0}^{\left( 1\right) }=\sqrt{\frac{a_{31}}{a_{20}}},  \label{d(1)0}
\end{equation}%
Hence the gapless branch, in the long-wavelength limit, is linear in $p$ and
represents the phonon mode with the speed of sound determined by Eq.~(\ref%
{d(1)0}). The dotted lines in Fig.~\ref{fig2} represent the
spectra with collisions in the stable regime. Once again, they are
calculated by solving the cubic Eq.~(\ref{eigenvalue equation}),
but are found in good agreement with the perturbative results (not
shown).

According to Eq.~(\ref{d(1)0}), whenever $a_{31}$ and $a_{20}$ are
opposite in sign, $\omega _{0}$ becomes imaginary for finite $p$
in the long wavelength limit, indicating an unstable 0-branch. We
now turn to a detailed discussion of the collision-induced
dynamical instability. Note that from Eq.~(\ref{d(1)0}), it may
seem that $ \omega _{0}$ is singular for any finite momenta when
$a_{20}=0$. However, when $a_{20}=0,\omega _{0}^{\left( 0\right)
}$ and $\omega _{-}^{\left( 0\right) }$ are degenerate according
to Eq.~(\ref{omega (0)}). Thus, this singularity is only an
artefact, indicating the failure of the nondegenerate perturbative
calculation. Indeed, exact solutions of Eq.~(\ref{eigenvalue
equation}) do not exhibit such a singularity. Despite of this
complication, the perturbative results can be used to identify the
unstable regimes quite accurately.

\section{Classification of Dynamical Instability}

\label{classification section}

Studies in Sec.~\ref{spectra} suggest that complex excitation frequencies
and hence dynamical instabilities can occur under the following situations:

\begin{enumerate}
\item When $a_{31}$ and $a_{20}$ possess opposite signs, 0-branch becomes
unstable [see Eq.~(\ref{d(1)0})];

\item When $a_{20}<0$, $\omega _{-}$ becomes complex [see Eq.~(\ref{omega
(0)})];

\item When $a_{10}^{2}<4a_{20}$, both $(\pm )$-branches become unstable [see
Eq.~(\ref{omega (0)})].
\end{enumerate}

Before moving foward, let us clarify our unit system. We consider
systems with particle densities in the order of $n_{0}=5\times
10^{20}$ m$ ^{-3}$, and $\alpha $ fixed to $4.22\times 10^{-6}$
m$^{3/2}$s$^{-1}$, which corresponds to the atom-molecule coupling
strength for the $^{23}$Na Feshbach resonance at a magnetic field
strength of $85.3$ mT \cite{seaman03}. We then adopt a unit system
in which $n_{0}$ is the unit for density, $ \alpha
/\sqrt{n_{0}}=1.88724\times 10^{-16}$m$^{3}$s$^{-1}$ the unit for
collisional parameters ($ \lambda _{i}$, $\lambda _{ij}$), $\alpha
\sqrt{n_{0}}=9.436\times 10^{4}$ s$ ^{-1}$ the unit for
frequencies $\left( \omega ,\Omega ,\epsilon \right) $, and
$\sqrt{ \left( 2m\hbar \alpha \sqrt{n_{0}}\right) }=8.72\times
10^{-28}$ kg m/s the unit for momentum $p$. Furthermore, for
simplicity, we only consider the situation where all the
scattering lengths are positive.

Our goal is to identify the unstable regimes in the parameter space spanned
by $\epsilon $ and $\Omega $, while the laser detuning $\Delta $ is always
fixed by the resonance condition represented by Eq.~(\ref{Delta arbitrary}).
For cases 1 and 2, we can identify two threshold values for the Feshbach
detuning $\epsilon _{0}^{th}$ and $\epsilon _{1}^{th}$, given by
\begin{eqnarray}
\epsilon _{0}^{th} &=&\Lambda _{a}\left\vert \psi _{a}^{0}\right\vert
^{2}+\Lambda _{g}\left\vert \psi _{g}^{0}\right\vert ^{2}+\frac{2\left(
\frac{\Omega ^{2}}{4|\psi _{a}^{0}|^{2}}+1\right) ^{2}}{4\lambda
_{a}+\lambda _{g}-4\lambda _{ag}},  \label{e 0} \\
\epsilon _{1}^{th} &=&\Lambda _{a}\left\vert \psi _{a}^{0}\right\vert
^{2}+\Lambda _{g}\left\vert \psi _{g}^{0}\right\vert ^{2}  \notag \\
&&\;+\frac{\lambda _{a}\Omega ^{2}+4\left( \lambda _{g}\left\vert \psi
_{g}^{0}\right\vert ^{2}+\lambda _{ag}\left\vert \psi _{a}^{0}\right\vert
^{2}\right) }{8\left\vert \psi _{g}^{0}\right\vert ^{2}\left( \lambda
_{a}\lambda _{g}-\lambda _{ag}^{2}\right) },  \label{e 1}
\end{eqnarray}%
which are obtained from the conditions $a_{20}=0$ and $ a_{31}=0$,
respectively. Case 3 will be discussed at the end of this section.

\begin{figure}[h]
\centering \includegraphics[width=2.5in]{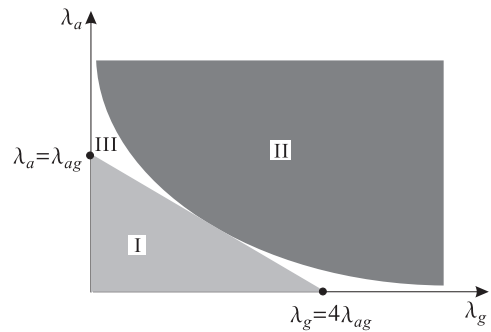}
\caption{Division of the positive quarter of $\left( \protect\lambda _{g},%
\protect\lambda _{a}\right) $-space into three regions by a straight line: $%
\protect\lambda _{g}=-4\protect\lambda _{a}+4\protect\lambda _{ag}$ and a
hyperbola: $\protect\lambda _{a}\protect\lambda _{g}=\protect\lambda %
_{ag}^{2}$.}
\label{fig3}
\end{figure}

It is not difficult to see that the signs of the denominators of
the last terms of (\ref{e 0}) and (\ref{e 1}), $4\lambda
_{a}+\lambda _{g}-4\lambda _{ag}$ and $ \lambda _{a}\lambda
_{g}-\lambda _{ag}^{2}$, respectively, will play crucial roles in
locating the instability parameter regime. For this reason, we
first decompose the $\left( \lambda _{g}\text{,}\lambda
_{a}\right) $-space into three regions divided by a straight line,
$\lambda _{g}=-4\lambda _{a}+4\lambda _{ag}$, and a hyperbola,
$\lambda _{a}\lambda _{g}=\lambda _{ag}^{2}$, as shown in
Fig.~\ref{fig3}. The line is tangent to the hyperbola at the point
$\left( \lambda _{g}=2\lambda _{ag}\text{,}\lambda _{a}=0.5\lambda
_{ag}\right) $. The three regions are defined as follows: Both
$4\lambda _{a}+\lambda _{g}-4\lambda _{ag}$ and $\lambda
_{a}\lambda _{g}-\lambda _{ag}^{2}$ are negative in Region I and
both are positive in Region II, while in Region III, $4\lambda
_{a}+\lambda _{g}-4\lambda _{ag}<0$ and $\lambda _{a}\lambda
_{g}-\lambda _{ag}^{2}>0$.

\begin{figure}[h]
\centering \includegraphics[width=2.5in]{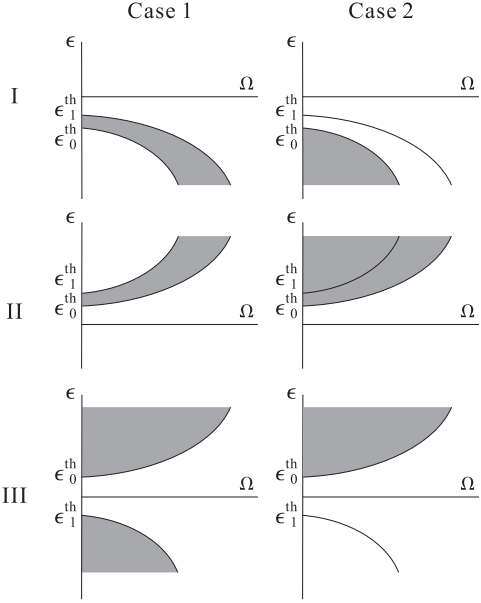}
\caption{Classification of the instabilities in the limit of low momentum.
The shaded areas of the diagrams in column one and two correspond to the
instabilities due to the imaginary sound speed and the complex gaps,
respectively. Each set of diagrams in row 1, 2 and 3 are the instability
maps for systems operating in region I, II, and III as specified in Fig. 3,
respectively.}
\label{fig4}
\end{figure}

Next, we choose to construct, for each region defined above, two
instability maps in the $\left( \Omega ,\epsilon \right) $-space;
one for the instability due to the imaginary sound speed (case 1),
and the other for the instability due to the complex gap (case 2).
We represent them by the shaded areas of the diagrams in the first
and second columns of Fig.~\ref{fig4}, respectively. Each set of
diagrams in row 1, 2 and 3 are built upon the collisional
parameters corresponding to Regions I, II, and III of $\left(
\lambda _{g}\text{,} \lambda _{a}\right) $-space, respectively. We
construct these maps based on two principles. First, it can be
shown from Eqs.~(\ref{e 0}) and (\ref{e 1}) that $\epsilon
_{1}^{th}-\epsilon _{0}^{th}\propto 1/\left( 4\lambda _{a}+\lambda
_{g}-4\lambda _{ag}\right) \left( \lambda _{a}\lambda _{g}-\lambda
_{ag}^{2}\right) $ with a positive definite proportionality. \ As
a result, we have $\epsilon _{1}^{th}>\epsilon _{0}^{th}$ in
Regions I and II, and $\epsilon _{1}^{th}<\epsilon _{0}^{th}$\ in
Region III. Second, according to Eq.~(\ref{e0 large Omega})
[Eq.~(\ref{e1 large Omega} )], $\epsilon _{0}^{th}$ [$\epsilon
_{1}^{th}$] is bent downward if $ 4\lambda _{a}+\lambda
_{g}-4\lambda _{ag}$ \thinspace $<0$ [$\lambda _{a}\lambda
_{g}-\lambda _{ag}^{2}$\thinspace $<0$] and vice versa. Note that
the instability of case 2 is solely determined by $\epsilon
_{0}^{th}$; the presence of $\epsilon _{1}^{th}$ in the second
column of Fig.~\ref{fig4} is to make the comparison of the two
maps easier. For example, one can easily identify that the two
instability maps in region II overlap in the band between
$\epsilon _{0}^{th}$ and $\epsilon _{1}^{th}$.

Let us now consider a specific atom-molecule system composed of $^{23}$Na.
The atomic $s$-wave scattering length for sodium is around $a_{a}=3.4$ nm
\cite{abeelen99} which yields $\lambda _{a}=0.625$. No precise knowledge
about the scattering lengths involving molecules exists. Assuming they are
on the same order of magnitude as $a_{a}$, we take $\lambda _{m}=\lambda
_{g}=\lambda _{am}=\lambda _{mg}=\lambda _{ag}=0.3\lambda _{a}$. This set of
parameters puts the system in Region II of Fig.~\ref{fig4}. According to
Eqs.~(\ref{e0 small Omega}) and (\ref{e1 small Omega}), $\epsilon _{0}^{th}$
and $\epsilon _{1}^{th}$ approach $0.5\Lambda _{g}+2\alpha ^{2}/\left(
4\lambda _{a}+\lambda _{g}-4\lambda _{ag}\right) =1.126$ and $0.5\Lambda
_{g}+0.5\alpha ^{2}\lambda _{g}/\left( \lambda _{a}\lambda _{g}-\lambda
_{ag}^{2}\right) =1.237,$ respectively, in the limit of small $\Omega $.

\begin{figure}[h]
\centering \includegraphics[width=2.5in]{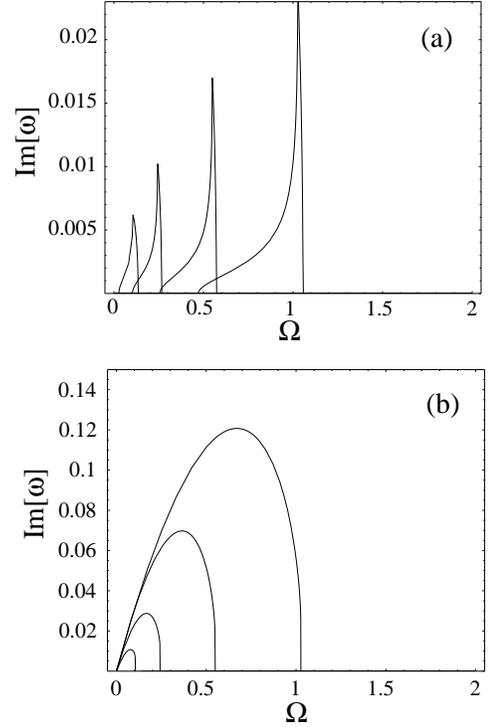}
\caption{Im$\left[ \protect\omega \left( p\right) \right] $ as a
function of $\Omega $ for a fixed momentum $p=0.005$. The curves
from left to right correspond to $\protect\epsilon =1.3$, $
\protect\epsilon =1.5$, $\protect\epsilon =2,$ and
$\protect\epsilon =3$, respectively.
Other parameters are $n=1$, $\protect\lambda _{a}=0.625,$ and $\protect%
\lambda _{m}=\protect\lambda _{g}=\protect\lambda _{am}=\protect\lambda %
_{mg}=\protect\lambda _{ag}=0.3\protect\lambda _{a}$. Units are defined in
Sec.~\protect\ref{classification section}.}
\label{fig5}
\end{figure}

A search using Eq.~(\ref{eigenvalue equation}) above these thresholds and at
small momenta indeed leads to two branches whose frequencies have finite
imaginary parts for certain values of $\Omega $, as illustrated in Fig.~\ref%
{fig5}(a) and (b), respectively. Each curve in Fig.~\ref{fig5}(a) is
restricted between two thresholds, both of which increase as $\epsilon $,
evidently resembling the instability originating from the imaginary sound
speed (case 1). Thus, in the limit of small $p$, we expect it to approach $%
d_{0}^{\left( 1\right) }p$. In fact, one can trace the sharp asymmetry of
each curve to the fact that according to Eq.~(\ref{d(1)0}), $d_{0}^{\left(
1\right) }$ vanishes at the left threshold where $a_{31}$ is close to zero,
while becomes singular at the right threshold where $a_{20}$ is close to
zero. In contrast, each curve in Fig.~\ref{fig5}(b) has only one threshold,
which matches quite closely to the right threshold of the corresponding
curve in Fig.~\ref{fig5}(a). This along with the fact that the value of the
mode frequency is rather insensitive with respect to $p$ (as long as $p$ is
not too large) indicates that it represents the instability from the complex
gapped branch of case 2. Therefore these exact calculations are in good
agreement with the insights gained from the instability map shown in Fig.~%
\ref{fig4}.

\begin{figure}[h]
\centering \includegraphics[width=2.5in]{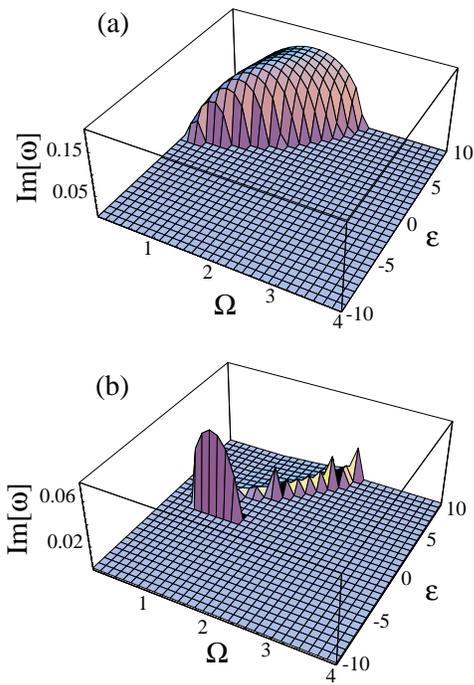} \caption{(Color
online) A three-dimensional view of Im$\left[ \protect\omega
\left( p\right) \right] $ as a function of $\Omega $ and
$\protect\epsilon $ for two branches of roots to
Eq.(\protect\ref{eigenvalue equation}). Parameters are same as in
Fig.~\protect\ref{fig5}. Units are defined in Sec.  \protect
\ref{classification section}.} \label{fig6}
\end{figure}

In order to have a more detailed look at various instabilities, we
make a three-dimensional plot in Fig.~\ref{fig6} for the imaginary
parts of the two (non-vanishing) branches of excitation
frequencies. A glance at this 3D view further confirms the
instability map of Fig.~4: one can easily associate the large bump
in Fig.~\ref{fig6}(a) to the instability of case 2, and the curved
band in Fig.~\ref{fig6}(b) to the instability of case 1. The
shapes of these unstable regions completely match with those shown
in Fig.~\ref{fig4}.

However, Fig.~\ref{fig6}(b) also reveals a new feature consisting
of a narrow strip along the $\Omega $ dimension. We can trace this
new feature to the source of instability associated with case 3
which has so far been ignored. An analysis with the help of
Eqs.~(\ref{a10}) and (\ref{a20}) indicates that the condition
$a_{20}>a_{10}^{2}/4$ amounts to
\begin{eqnarray}
&&\epsilon ^{\prime }\left[ \epsilon ^{\prime 3}+\left( \Omega
^{2}+4\alpha ^{2}\left\vert \psi _{a}^{0}\right\vert ^{2}\right)
\epsilon ^{\prime } \right. \nonumber \\ &&\left.+2\alpha
^{2}\left\vert \psi _{a}^{0}\right\vert ^{4}\left( \lambda
_{g}-4\lambda _{ag}+4\lambda _{a}\right) \right] <0, \label{new
instability}
\end{eqnarray}%
from which we find that to a good approximation, the unstable
$\epsilon $ is bounded between
\begin{subequations}
\begin{eqnarray}
\epsilon_1 &=& \Lambda _{a}\left\vert \psi _{a}^{0}\right\vert
^{2}+\Lambda _{g}\left\vert \psi _{g}^{0}\right\vert ^{2}\,, \\
\epsilon_2 & \approx &\epsilon_1-2\alpha ^{2}\left\vert \psi
_{a}^{0}\right\vert ^{4}\,\frac{ \lambda _{g}-4\lambda
_{ag}+4\lambda _{a}}{ \Omega ^{2}+4\alpha ^{2}\left\vert \psi
_{a}^{0}\right\vert ^{2}}\,,
\end{eqnarray}
\end{subequations}
It can be shown that $\epsilon _1$ and $\epsilon _2$
are not very sensitive to $\Omega $, and the they merge together
in the limit of either small or large $\Omega $, which explains
the shape of this new instability region.

\section{Dynamical Signatures of Stable and Unstable CPT states}

In order to determine the fate of an unstable state, we must
examine its dynamical response to fluctuations, which can be
simulated using the time-dependent version of the GPEs, which are
obtained by replacing the left hand sides of Eqs.~(\ref{steady
state equation}) by $i \dot{\psi}_i$.

We take a CPT state associated with a Rabi frequency $\Omega _{0}$
and Feshbach detuning $\epsilon $ to be the initial state. To
simulate small fluctuations, we modify the Rabi frequency as
\begin{equation}
\Omega =\Omega _{0}+\left[ \delta \Omega _{0}+2\delta \Omega _{1}\sin \left(
px/\hbar \right) \right] ,  \label{Omega standing wave}
\end{equation}%
where $\delta \Omega _{0}$ and $\delta \Omega _{1}$ are small
perturbation amplitudes. Note that a position-dependent
perturbation represented by the $ \delta \Omega_1$ term can cause
spatial deformation of the uniform CPT state, making it possible
to trigger the instabilities due to perturbations of finite
momentum.

\begin{figure}[h]
\centering \includegraphics[width=2.5in]{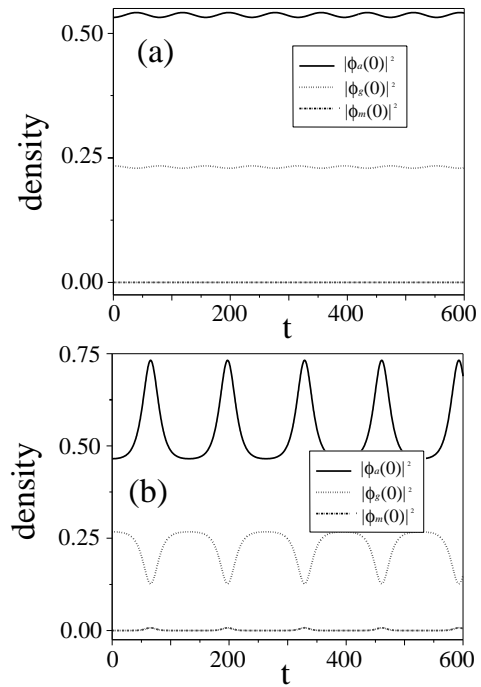} \caption{The
time evolution of the particle number densities in the condensate
mode probed by a zero-momentum perturbation with $\protect\delta
\Omega _{0}=0.001$ for (a) $\Omega _{0}=1.1$ and (b) $\Omega
=0.9$, while the Feshbach detuning is fixed at $\protect\epsilon
=3$. Other parameters are the same as in Fig.~\protect\ref{fig5}.
Units are defined in Sec.~\protect\ref{classification section}.}
\label{fig7}
\end{figure}
To test the instability of case 2, we take the same collisional parameters
as in Fig.~\ref{fig5}, but fix the Feshbach detuning to $\epsilon =3$. We
apply a zero-momentum perturbation with an amplitude $\delta \Omega
_{0}=0.001$ and $\delta \Omega _{1}=0$ to the CPT state with (a) $\Omega
_{0}=1.1$ and (b) $\Omega _{0}=0.9$, respectively. By inspecting the $%
\epsilon =3$ curve of Fig.~\ref{fig5}(b), which survives even when
$p=0$, we anticipate that the dynamics of the system will change
from a stable to an unstable one as $\Omega _{0}$ varies from
$1.1$ to $0.9$. The results, shown in Fig.~\ref{fig7}(a) and (b),
clearly support our analysis. The stable CPT state does not
respond significantly to the small perturbation; while
Fig.~\ref{fig7}(b) shows a large-amplitude self-pulsing behavior,
characteristic of an unstable system. In addition, there is a
noticeable population in the unstable molecular state. Such
self-pulsing oscillations are also familiar features in the study
of unstable nonlinear optical systems \cite{pierre}.

\begin{figure}[h]
\centering \includegraphics[width=2.8in]{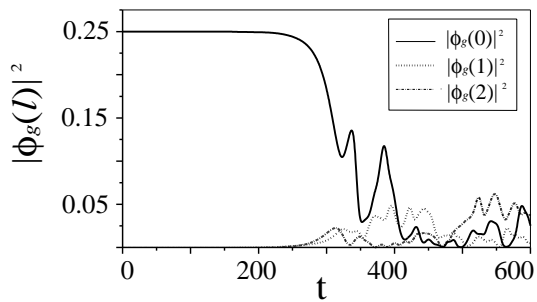} \caption{The
time evolution of $\left\vert \protect\phi _{g}\left( l,t\right)
\right\vert ^{2}$ ($l=0,1,$ and $2$)\ after a finite-momentum
perturbation with $\protect\delta \Omega _{1}=0.001$ at $p=0.01$
$\left( \protect\delta \Omega _{0}=0\right) $ is applied to a
system initially prepared in a CPT\ state of $\protect\epsilon
=-4$ and $\Omega _{0}=1$. We include modes ranging from
$l=-10$ to 10 in the simulation. Other parameters are $\protect%
\lambda _{a}=0.625$, $\protect\lambda _{m}=\protect\lambda
_{g}=\protect \lambda _{mg}=0.5\protect\lambda _{a}$,
$\protect\lambda _{am}=\protect \lambda _{a}$, $\protect\lambda
_{ag}=1.4\protect\lambda _{a},$ and $n=1$. Units are defined in
Sec.~\protect\ref{classification section}.} \label{fig8}
\end{figure}

To demonstrate the instability of case 1, we need to apply a
finite-momentum perturbation since for case 1, the imaginary part
of the excitation frequency $\omega _{0}$ vanishes at $p=0$. The
finite-momentum perturbation will couple together different
momentum modes. To simulate this process, we adopt the Floquet
technique and expand the wave functions as
\begin{equation}
\psi _{i}\left( \mathbf{r},t\right) =\sum_{l=-\infty }^{+\infty
}\phi _{i}\left( l,t\right) e^{ilpx/\hbar }\,,
\end{equation}%
where $\left\vert \phi _{i}\left( l,t\right) \right\vert ^{2}$ represent the
particle density of the $i$th species with momentum $lp$. \ Here, $l$ is an
integer and $p=2\pi \hbar /L$, with $L$ being the length of the condensate
along $x$ dimension. Then the orthonormality condition for the momentum
modes reads
\begin{equation}
\frac{1}{L}\,\int e^{i\left( l-l^{\prime }\right) px/\hbar
}dx=\delta _{l,l^{\prime }}\,.
\end{equation}
This enables us to transform the dynamical GPEs for $\psi _{i}\left( \mathbf{%
r},t\right) $ into the coupled modal equations for $\phi _{i}\left(
l,t\right) $
\begin{widetext}
\begin{subequations}
\begin{eqnarray}
i\frac{d\phi _{a}\left( n\right) }{dt} &=&-\epsilon_p n^{2}\phi
_{a}\left( n\right) +\sum_{l,l^{\prime }}\sum_{s}\lambda _{as}\phi
_{s}\left( l\right) \phi _{s}^{\ast }\left( l^{\prime }\right)
\phi _{a}\left( n+l^{\prime }-l\right) +\alpha \sum_{l}\phi
_{m}\left( l\right) \phi _{a}^{\ast }\left(
l-n\right) \,, \\
i\frac{d\phi _{m}\left( n\right) }{dt}
&=&-\frac{\epsilon_p}{2}n^{2}\phi _{m}\left( n\right)
+\sum_{l,l^{\prime }}\sum_{s}\lambda _{ms}\phi _{s}\left( l\right)
\phi _{s}^{\ast }\left( l^{\prime }\right) \phi
_{m}\left( n+l^{\prime }-l\right) +\epsilon \phi _{m}\left( n\right) \nonumber \\
&&+\frac{\alpha }{2}\sum_{l}\phi _{a}\left( l\right) \phi
_{a}\left( n-l\right) -\frac{\Omega _{0}+\delta \Omega
_{0}}{2}\phi _{g}\left( n\right) -\frac{\delta \Omega
_{1}}{2i}\left[ \phi _{g}\left( n-1\right) -\phi
_{g}\left( n+1\right) \right]\, , \\
i\frac{d\phi _{g}\left( n\right) }{dt}
&=&-\frac{\epsilon_p}{2}n^{2}\phi _{g}\left( n\right)
+\sum_{l,l^{\prime }}\sum_{s}\lambda _{gs}\phi _{s}\left( l\right)
\phi _{s}^{\ast }\left( l^{\prime }\right) \phi _{g}\left(
n+l^{\prime }-l\right) +\left( \Delta +\epsilon \right) \phi
_{g}\left( n\right) \nonumber \\
&&-\frac{\Omega _{0}+\delta \Omega _{0}}{2}\phi _{m}\left( n\right) -\frac{%
\delta \Omega _{1}}{2i}\left[ \phi _{m}\left( n-1\right) -\phi
_{m}\left( n+1\right) \right]\, .
\end{eqnarray}%
\end{subequations}
\end{widetext}
The initial condition is then equivalent to
\begin{equation}
\phi _{a}\left( 0,0\right) =\left\vert \psi _{a}^{0}\right\vert
,\phi _{m}\left( 0,0\right) =0,\phi _{g}\left( 0,0\right)
=\left\vert \psi _{g}^{0}\right\vert \,,
\end{equation}%
and $\phi _{i}\left( l\neq 0,0\right) =0$, where $\left\vert \psi
_{a}^{0}\right\vert $ and $\left\vert \psi _{g}^{0}\right\vert $
are determined from Eqs.~(\ref{CPT distribution}) with $\Omega $
being replaced by $\Omega _{0}$. We now switch to a set of
collisional parameters with $\lambda _{a}=0.625$, $\lambda
_{m}=\lambda _{g}=\lambda _{mg}=0.5\lambda _{a}$, $\lambda
_{am}=\lambda _{a}$, and $\lambda _{ag}=1.4\lambda _{a}$, which
puts the system in Region I, where the two instabilities regions
of case 1 and 2 do not overlap in the same parameter space (see
the two diagrams in the first row of Fig.~\ref{fig4}). We take
$\epsilon =-4$ and $ \Omega _{0}=1$, which puts the system in the
unstable region of case 1. This allows us to demonstrate the
instability due solely to the finite-momentum perturbation.
Indeed, simulations show that the system is stable against
zero-momentum perturbations but is unstable when finite-momentum
perturbations are applied. A dynamical result of a finite-momentum
perturbation is displayed in Fig.~\ref{fig8} which shows that the
instability leads to an irreversible population transfer from the
condensate mode ($p=0$) to non-condensate modes ($p\neq 0$). The
initial growth of the non-condensate modes is approximately
exponential but quickly becomes rather chaotic, which is typical
for an unstable multi-mode system \cite{ling01}.

\section{Conclusion}

To summarize, using a generalized Bogoliubov approach, we have
calculated the collective excitation spectra of the CPT states
involving atomic and stable molecular condensates. We have found,
through our analysis of the excitation spectra, that collisions
involving atoms and stable molecules strongly affect the stability
properties of the CPT state. We have developed a set of analytical
criteria for classification of various instabilities in the
long-wavelength limit. We have shown that these criteria can be
used to accurately identify the unstable parameter regimes. In
this paper, we studied a homogeneous system. It will be desirable
in the future to extend this work to the inhomogeneous case where
trapping potentials are present, as in typical experimental
situations. Nevertheless, our work here should provide important
analytical insights into the much more complicated inhomogeneous
problem, for which the only viable treatment is perhaps through
numerical simulations.

Just as the usual CPT states in linear three-level $\Lambda
$-systems are behind many important applications in quantum and
atom optics \cite{arimondo96}, we anticipate that the nonlinear
CPT states will also play important roles in applications where
the coherence between atom and molecule condensates becomes
critical. Evidently, the success of these applications depends on
our ability to avoid the unstable regimes of the CPT state. For
example, high efficiency coherent population transfer between the
atomic and molecular condensate using STIRAP can only be achieved
if these unstable regimes are not encountered during the transfer
process \cite{ling04}. With this in mind, we believe that our work
will have important implications in the coupled atom-molecule
condensates \cite{heinzen00}, a system of significant current
interest, with ramifications in other nonlinear systems.

\bigskip

\section{Acknowledgments}

HYL acknowledges the support from the US National Science
Foundation under Grant No. PHY-0307359. We thank Dr. Su Yi for
many discussions and help with some of the figures.

\bigskip
\begin{appendix}%

\section{Asymptotes}

A qualitative understanding of the instability map shown in Fig.~4 depends
on the asymptotic behavior of the threshold Feshbach detunings $%
\epsilon_0^{th}$ and $\epsilon_1^{th}$ which we show below. In the limit of
large $\Omega /\alpha $
\begin{eqnarray}
\epsilon _{0}^{th} &\approx &\Lambda _{a}+\frac{\alpha ^{2}}{8\left(
4\lambda _{a}+\lambda _{g}-4\lambda _{ag}\right) }\left( \frac{\Omega }{%
\alpha }\right) ^{4},  \label{e0 large Omega} \\
\epsilon _{1}^{th} &\approx &\Lambda _{a}+\frac{\lambda _{a}\alpha ^{2}}{%
8\left( \lambda _{a}\lambda _{g}-\lambda _{ag}^{2}\right) }\left( \frac{%
\Omega }{\alpha }\right) ^{4},  \label{e1 large Omega}
\end{eqnarray}
and in the limit of small $\Omega /\alpha $
\begin{widetext}
\begin{eqnarray}
\epsilon _{0}^{th} &\approx &0.5\Lambda _{g}+\frac{2\alpha ^{2}}{4\lambda
_{a}+\lambda _{g}-4\lambda _{ag}}+\left( \frac{2\Lambda _{a}-\Lambda_{g}}{2\sqrt{2}}+\frac{%
\sqrt{2}\alpha ^{2}}{4\lambda _{a}+\lambda _{g}-4\lambda _{ag}}\right) \frac{\Omega }{\alpha },
\label{e0 small Omega} \\
\epsilon _{1}^{th} &\approx &0.5\Lambda _{g}+\frac{0.5\alpha ^{2}\lambda _{g}}{%
\lambda _{a}\lambda _{g}-\lambda _{ag}^{2}}+\frac{\lambda _{ag}^{2}\left(
2\lambda _{ag}-\Lambda _{g}\right) +\lambda
_{a}\lambda _{g}\left(\Lambda _{g}-2\Lambda
_{a}\right) -2\alpha ^{2}\lambda _{ag}}{2\sqrt{2}\left( \lambda
_{ag}^{2}-\lambda _{a}\lambda _{g}\right) }\frac{\Omega }{\alpha },
\label{e1 small Omega}
\end{eqnarray}
\end{widetext}

\end{appendix}%

\setcounter{secnumdepth}{-1}%

%

\bigskip


\begin{thebibliography}{1}


\bibitem{Feshbach}
H. Feshbach, {\em Theoretical Nuclear Physics} (Wiley, New York,
1992); E. Tiesinga, A. J. Moerdijk, B. J. Verhaar, and H. T. C.
Stoof, Phys. Rev A {\bf 46}, R1167 (1992).

\bibitem{timmermans99}
E. Timmermans, P.
Tommasini, M. Hussein, and A. Kerman, Phys. Rep. {\bf 315}, 199
(1999).

\bibitem{donley02}
E. A. Donley {\em et al.}, Nature (London) {\bf 417}, 529 (2002);
J. Herbig {\em et al.}, Science {\bf 301}, 1510 (2003); K. Xu {\em
et al.}, Phys. Rev. Lett. {\bf 91}, 210402 (2003); S. D\"{u}rr
{\em et al.}, Phys. Rev. Lett. {\bf 92}, 020406 (2004).


\bibitem{drummond98}
P. D. Drummond and K. V. Kheruntsyan, Phys. Rev. Lett. {\bf 15}, 3055 (1998).

\bibitem{javanainen99}
J. Javanainen and M. Mackie, Phys. Rev. A {\bf 59}, 3186(R) (1999).

\bibitem{yurovsky99}
V. A. Yurovsky, A. Ben-Reuven, P. S. Julienne, and C. J. Williams, Phys. Rev. A {\bf 60}, 765 (1999).

\bibitem{heinzen00} D. J. Heinzen, R. Wynar, P. D. Drummond,
and K. V. Kheruntsyan, Phys. Rev. Lett. {\bf 84}, 5029 (2000).

\bibitem{kokkelmans02} S. J. J. M. F. Kokkelmans and M. J. Holland, Phys. Rev. Lett. {\bf 89},
180401 (2002).

\bibitem{javanainen02}J. Javanainen and M. Mackie, Phys. Rev. Lett. {\bf 88}, 090403 (2002).

\bibitem{duine04}
R. A. Duine and H. C. Stoof, Phys. Repts. {\bf 396}, 115 (2004).

\bibitem{fbecnote} Several groups have successfully created molecular
condensate by magneto-associating degenerate Fermi atoms [M.
Greiner, C. Regal and D. S. Jin, Nature {\bf 426}, 537 (2003); S.
Jochim {\em et al.}, Science {\bf 302}, 2101 (2003); M. W.
Zwierlein {\em et al.}, Phys. Rev. Lett. {\bf 91}, 250401 (2003);
T. Bourdel {\em et al.}, Phys. Rev. Lett. {\bf 93}, 050401
(2004).]. Although the resulting molecules are still in high
vibrational levels, their decay into atomic constituents are
suppressed by Pauli blocking, giving rise to long lifetimes for
the molecular condensate [D. S. Petrov, C. Salomon, and G. V.
Shlyapnikov, Phys. Rev. Lett. {\bf 93}, 090404 (2004)].

\bibitem{thermal}A. Vardi {\em et al.}, J. Chem. Phys. {\bf107},
6166 (1997).
\bibitem{Mackie00}
M. Mackie, R. Kowalski, and J. Javanainen, Phys. Rew. Lett.
{\bf84}, 2000; M. Mackie {\em et al.}, Phys. Rev. A {\bf 70},
013614 (2004).


\bibitem{hope01}
J. J. Hope, M. K. Olsen, and L. I. Plimak, Phys. Rev. A {\bf 63}, 043603 (2001).

\bibitem{Drummond02}
P. D. Drummond, K. V. Kheruntsyan, D. J. Heinzen, and R. H. Wynar,
Phys. Rev. A {\bf 65}, 063619 (2002).

\bibitem{Hioe83} F. T. Hioe, Phys. Lett. A {\bf 99}, 150 (1983); F. T. Hioe and J. H. Eberly,
Phys. Rev. A {\bf 29}, 1164(1984); J. Oreg, F. T. Hioe, and J. H.
Eberly, Phys. Rev. A {\bf 29}, 690 (1984).

\bibitem{stirap}K. Bergmann, H. Theuer and B. W. Shore, Rev. Mod.
Phys. {\bf70}, 1003 (1998).


\bibitem{Gray78}   H. R. Gray, R. M. Whitley, and C. R. Stroud, Jr., Opt. Lett. {\bf 3}, 218 (1978).
\bibitem{Alzetta76}
G. Alzetta, A. Gozzini, L. Moi and G. Orriols, Nuovo Cimento B
{\bf 36}, 5 (1976); G. Alzetta, L. Moi and G. Orriols, Nuovo
Cimento B {\bf52}, 209 (1979).


\bibitem{kokk}S.J.J.M.F. Kokkelmans, H.M.J. Vissers and B. J.
Verhaar, Phys. Rev. A {\bf63}, 031601(R) (2001).
\bibitem{mackie} M. Mackie, Phys. Rev. A {\bf66}, 043613 (2002).

\bibitem{Harris02}
S. E. Harris, Phys. Rev. A {\bf 66}, 010701(R) (2002).

\bibitem{ling04}H. Y. Ling, H. Pu, and B. Seaman, Phys. Rev. Lett.
{\bf 93}, 250403 (2004)




\bibitem{bogoliubov47}
N. N. Bogoliubov, J. Phys. USSR {\bf 11}, 23 (1947).

\bibitem{fetter71}
A. L. Fetter and J. D. Walecka, Quantumm Theory of Many-Particle Systems,
(McGraw-Hill, New York, 1971).

\bibitem{comment}
There are six roots. However, for the stability analysis, it
suffices to examine the three positive branches defined by
$\sqrt{\omega ^{2}}$, where $\omega ^{2}$ is the cubic solution of
the characteristic equation.

\bibitem{seaman03}
B. Seaman and H. Ling, Opt. Commun. {\bf226}, 267 (2003).

\bibitem{abeelen99}
F. A. van Abeelen, B. J. Verhaar, Phys. Rev. A {\bf59}, 578
(1999).

\bibitem{pierre} P. Meystre and M. Sargent III, {\em Elements of
Quantum Optics} (Springer-Verlag, Berlin, Heidelberg, 1999).

\bibitem{ling01}
H. Y. Ling, H. Pu, and L. Baksmaty and N. P. Bigelow, Phys. Rev. A
{\bf 63}, 053810 (2001); H. Y. Ling, Phys. Rev. A {\bf 63}, 053810 (2001).

\bibitem{arimondo96}
E. Arimondo, Progress in Optics {\bf 35}, 257 (1996).




\end{thebibliography}
\end{document}